\documentclass[12pt]{article}
\usepackage{graphics}
\usepackage{amssymb}

\newcommand{\QED}{\mbox{\rule[-1.5pt]{6pt}{10pt}}}
\newtheorem{claim}{Claim}[section]
\newtheorem{theorem}[claim]{Theorem}
\newtheorem{proposition}[claim]{Proposition}
\newtheorem{lemma}[claim]{Lemma}
\newtheorem{corollary}[claim]{Corollary}
\begin{document}

\title{Persistent currents for 2D Schr{\"o}dinger operator
 with a strong $\delta$-interaction on a loop}
\author{P.~Exner$^{a,b}$ and K.~Yoshitomi$^{c}$}
\date{}
\maketitle
\begin{quote}
{\small \em a) Department of Theoretical Physics, Nuclear Physics
Institute, \\ \phantom{e)x}Academy of Sciences, 25068 \v Re\v z,
Czech Republic \\
 b) Doppler Institute, Czech Technical University,
 B\v{r}ehov{\'a} 7,\\
\phantom{e)x}11519 Prague, Czech Republic \\
 c) Graduate School of Mathematics, Kyushu University,
Hakozaki, \\ \phantom{e)x}Fukuoka 812-8581, Japan \\
 \rm \phantom{e)x}exner@ujf.cas.cz,
 yositomi@math.kyushu-u.ac.jp}
\vspace{8mm}

\noindent {\small We investigate the two-dimensional magnetic
 Schr\"odinger operator $H_{B,\beta}=\left(-i\nabla-A\right)^2
-\beta\delta(\cdot-\Gamma)$, where $\Gamma$ is a smooth loop and
 the vector potential $A$ corresponds to a homogeneous
 magnetic field $B$
 perpendicular to the plane. The asymptotics of negative
 eigenvalues of $H_{B,\beta}$ for $\beta\to\infty$ is found. It shows, in
 particular, that for large enough positive $\beta$ the system
 exhibits persistent currents.}
\end{quote}

%%%%%%%%%%%%%%%%%%%%%%%%%%%%%%%%%%%%%%%%%%%%%%%%%%%%%%%%%%%%%%%%%%%

\section{Introduction}

One of the most often studied features of mesoscopic systems are
 the persistent currents in rings threaded by a magnetic flux --
 let us mention, e.g., \cite{CGR, CWB} and scores of other
 theoretical and experimental papers where they were discussed.
 For a charged particle (an electron) confined to a loop $\Gamma$
 the effect is manifested by the dependence of the corresponding
 eigenvalues $\lambda_n$ on the flux $\phi$ threading the loop,
 conventionally measured in the units of flux quanta, $2\pi\hbar
 c|e|^{-1}$. The derivative $\partial \lambda_n/\partial \phi$
 equals $-{1\over c}I_n$, where $I_n$ is the persistent current
 in the $n$--th state. In particular, if the particle motion on
 the loop is free, we have
 % ------------- %%
 \begin{equation} \label{ideal}
\lambda_n(\phi) = {\hbar^2\over 2m^*} \left( 2\pi\over
L\right)^2 (n+\phi)^2,
 \end{equation}
 % ------------- %%
where $L$ is the loop circumference and $m^*$ is the effective
 mass of the electron, so the currents depend linearly on the
 applied field in this case.

Of course, the above example is idealized assuming that the
 particle is strictly confined to the loop. In reality boundaries
 of a quantum wire are potential jumps at interfaces of different
 materials. As a consequence, electrons can be found outside the
 loop, even if not too far when we consider energies at which the
 exterior is a classically forbidden region.

A reasonable model respecting the essentially one-dimensional
 nature of quantum wires is a 2D Schr\"odinger operator with an
 attractive $\delta$-interaction on an appropriate curve
 $\Gamma$, or more generally, a planar graph. Since the
 interaction support has codimension one, the Hamiltonian can be
 defined through its quadratic form and the corresponding
 resolvent can be written explicitly as a generalization of the
 Birman-Schwinger theory \cite{BT, BEKS}. This leads to some
 interesting consequences such as the existence of bound states
 due to bending of an infinite and asymptotically straight curve
 \cite{EI}.

A natural question is how such a model is related to the ideal
 one in which the electron is strictly confined to the curve
 $\Gamma$. In \cite{EY} we have derived an asymptotic formula
 showing that if the $\delta$ coupling is strong, the negative
 eigenvalues approach those of the ideal model in which the
 geometry of $\Gamma$ is taken into account by means of an
 effective curvature-induced potential. The purpose of this paper
 is to ask a similar question in the situation when the electron
 is subject in addition to a homogeneous magnetic field $B$
 perpendicular to the plane. We are going to derive an analogous
 asymptotic formula where now the presence of the magnetic field
 is taken into account via the boundary conditions specifying the
 domain of the comparison operator.

An easy consequence of this result is that for a strong enough
 $\delta$-interaction the negative eigenvalues of our Hamiltonian
 are not constant as functions of $B$, i.e. that the system
 exhibits persistent currents. Their further properties depend,
 of course, on the specific shape of $\Gamma$; this fact as well
 as the stability of such currents with respect to a disorder
 raise questions about optimal ways of interpreting the
 corresponding magnetic transport. We comment on this point in
 the concluding remarks.

%%%%%%%%%%%%%%%%%%%%%%%%%%%%%%%%%%%%%%%%%%%%%%%%%%%%%%%%%%%%%%%%%%%

\setcounter{equation}{0}
\section{Description of the model and the results}

As we have said above we are going to study the Schr{\"o}dinger
 operator in $L^2(\mathbb{R}^2)$ with a constant magnetic field
 and an attractive $\delta$-interaction on a loop. For the sake
 of simplicity we emply rational units, $\hbar =c= 2m^* =1$ and
 absorb the electron charge into the field intensity $B$. We
 shall use the circular gauge, $A(x,y)=\left(-{1\over
2}By,{1\over 2}Bx \right)$.

Let $\Gamma:[0,L]\owns s\mapsto (\Gamma_{1}(s),\Gamma_{2}(s))\in
\mathbb{R}^{2}$ be a closed counter-clockwise $C^{4}$ Jordan
 curve which is parametrized by its arc length. Given $\beta>0$
 and $B\in\mathbb{R}$, we define
 % ------------- %%
 $$ %\begin{equation} \label{}
 q_{B,\beta}[f]=\left\| \left(-i\partial_x +{1\over 2}By \right)f
 \right\|^{2}
 +\left\|\left(-i\partial_y -{1\over 2}Bx\right)f\right\|^{2} -\beta
 \int_{\Gamma}|f(x)|^{2}\,ds
 $$ %\end{equation}
 % ------------- %%
with the domain $H^{1}(\mathbb{R}^{2})$, where $\partial_x
\equiv \partial/\partial x$ etc., and the norm refers to
 $L^2(\mathbb{R}^2)$. It is straightforward to check that the
 form $q_{B,\beta}$ is closed and below bounded. We denote by
 $H_{B,\beta}$ the self-adjoint operator associated with it which
 can be formally written as
 % ------------- %%
 $$ %\begin{equation} \label{}
 H_{B,\beta}=\left(-i\nabla-A\right)^2
 -\beta\delta(\cdot-\Gamma)\,.
 $$ %\end{equation}
 % ------------- %%
Our main aim is to study the asymptotic behavior of the
 negative eigenvalues of $H_{B,\beta}$ as $\beta\to +\infty$.

Let $\gamma:[0,L]\owns s\mapsto
(\Gamma_{1}^{\prime\prime}\Gamma_{2}^{\prime}
-\Gamma_{2}^{\prime\prime}\Gamma_{1}^{\prime})(s) \in\mathbb{R}$
 be the signed curvature of $\Gamma$. We denote by $\Omega$ the
 region enclosed by $\Gamma$, with the area $|\Omega|$, and
 define the operator
 % ------------- %%
 $$ %\begin{equation} \label{}
 S_{B}=-\frac{d^{2}}{ds^{2}}-\frac{1}{4}\gamma(s)^{2}
 $$ %\end{equation}
 % ------------- %%
on $L^{2}((0,L))$ with the domain
 % ------------- %%
 $$ %\begin{equation} \label{}
 P_{B}=\{\,\varphi\in H^{2}((0,L));\quad
 \varphi^{(k)}(L)=\exp(-iB|\Omega|)\varphi^{(k)}(0),\;
 k=0,1\,\}\,,
 $$ %\end{equation}
 % ------------- %%
where $\varphi^{(k)}$ stands for the $k$-th derivative.

We fix $j\in\mathbb{N}$ and denote by $\mu_{j}(B)$ the $j$-th
eigenvalue of $S_{B}$ counted with multiplicity. Our main
results is the following claim.
 % ------------- %
 \begin{theorem} \label{main}
 Let $n$ be an arbitrary integer
 and let $\emptyset\neq I\subset\mathbb{R}$
 be a compact interval. Then there exists
 $\beta(n,I)>0$ such that
 % ------------- %%
 $$ %\begin{equation} \label{}
 \sharp\{ \sigma_{d}(H_{B,\beta})\cap (-\infty,0)\}\geq n
 \quad\mathrm{for}\quad \beta\geq \beta(n,I) \quad\mathrm{and}\quad
 B\in I\,.
 $$ %\end{equation}
 % ------------- %%
 For $\beta\geq \beta(n)$ and $B\in I$ we denote by
 $\lambda_{n}(B,\beta)$ the $n$-th eigenvalue of $H_{B,\beta}$
 counted with multiplicity. Then $\lambda_{n}(B,\beta)$ admits an
 asymptotic expansion of the form
 % ------------- %%
 $$ %\begin{equation} \label{}
 \lambda_{n}(B,\beta)= -\frac{1}{4}\beta^{2}+\mu_{n}(B)
 +\mathcal{O}(\beta^{-1}\ln\beta)\quad\mathrm{as}\quad
 \beta\to+\infty\,,
 $$ %\end{equation}
 % ------------- %%
 where the error term is uniform with respect to $B\in I$.
 \end{theorem}
 % ------------- %
Recall that the flux $\phi$ through the loop is $B|\Omega|/2\pi$
 in our units. The existence of persistent currents is then given
 by the following consequence of the above result.
 % ------------- %
 \begin{corollary} \label{nonconst}
 Let $\emptyset\neq I\subset\mathbb{R}$
 be a compact interval and let $n\in\mathbb{N}$. Then there
 exists a constant $\beta_{1}(n,I)>0$ such that the function
 $\lambda_{n}(\cdot,\beta)$ is not constant for
 $\beta\geq \beta_{1}(n,I)$.
 \end{corollary}
 % ------------- %

%%%%%%%%%%%%%%%%%%%%%%%%%%%%%%%%%%%%%%%%%%%%%%%%%%%%%%%%%%%%%%%%%%%

\setcounter{equation}{0}
\section{The proofs}

Since the spectral properties of $H_{B,\beta}$ are clearly
 invariant with respect to Euclidean transformation of the plane, we may
 assume without any loss of generality that the curve $\Gamma$
 parametrizes in the following way,
 % ------------- %%
 $$ %\begin{equation} %\label{}
 \Gamma_{1}(s) = \Gamma_{1}(0)+\int^{s}_{0}\cos
 H(t)\,dt\,, \quad
 \Gamma_{2}(s) = \Gamma_{2}(0)+\int^{s}_{0}\sin H(t)\,dt\,,
 $$ %\end{equation}
 % ------------- %%
where $H(t):=-\int^{t}_{0}\gamma(u)\,du $. Let $\Psi_{a}$ be the
 map
 % ------------- %%
 $$ %\begin{equation} %\label{}
 \Psi_{a}:\, [0,L)\times (-a,a)\owns (s,u)\mapsto
 (\Gamma_{1}(s)\!-\!u\Gamma_{2}^{\prime}(s),
 \Gamma_{2}(s)\!+\!u\Gamma^{\prime}_{1}(s))\in\mathbb{R}^{2}.
 $$ %\end{equation}
 % ------------- %%
By \cite[Lemma 2.1]{EY} we know that there exists $a_{1}>0$ such
 that the map $\Psi_{a}$ is injective for all $a\in (0,a_{1}]$.
 We fix thus $a\in (0,a_{1})$ and denote by $\Sigma_{a}$ the
 strip of width $2a$ enclosing $\Gamma$
 % ------------- %%
 $$ %\begin{equation} %\label{}
 \Sigma_{a}:=\Psi_{a}\left([0,L)\times (-a,a)\right)\,.
 $$ %\end{equation}
 % ------------- %%
Then the set $\mathbb{R}^{2}\backslash\Sigma_{a}$ consists of
 two connected components which we denote by
 $\Lambda^{\mathrm{in}}_{a}$ and $\Lambda^{\mathrm{out}}_{a}$,
 where the interior one, $\Lambda^{\mathrm{in}}_{a}$, is compact.
 We define a pair of quadratic forms,
 % ------------- %%
 \begin{eqnarray*} %\label{}
 \lefteqn{ q_{B,a,\beta}^{\pm}[f]=\left\Vert \left(-i\partial_x
 +{1\over 2}By \right)f \right\Vert^{2}_{L^{2}(\Sigma_{a})}
 +\left\Vert \left(-i\partial_y-{1\over 2}Bx \right)f
 \right\Vert^{2}_{L^{2}(\Sigma_{a})} } \\ &&
 \phantom{AAAA} -\beta\int_{\Gamma}|f(x)|^{2}\,ds \,,
 \phantom{AAAAAAAAAAAAAAAAAAAA}
 \end{eqnarray*}
 % ------------- %%
which are given by the same expression but differ by their
 domains; the latter is $H_0^{1}(\Sigma_{a})$ for
 $q_{B,a,\beta}^{+}$ and $H^{1}(\Sigma_{a})$ for
 $q_{B,a,\beta}^{-}$. Furthermore, we introduce the quadratic
 forms
 % ------------- %%
 $$ %\begin{equation} \label{}
 e_{B,a}^{j,\pm}[f]=\left\Vert \left(-i\partial_x
 +{1\over 2}By \right)f \right\Vert^{2}_{L^{2}(\Lambda^j_{a})}
 +\left\Vert \left(-i\partial_y-{1\over 2}Bx \right)f
 \right\Vert^{2}_{L^{2}(\Lambda^j_{a})}
 $$ %\end{equation}
 % ------------- %%
for $j=\mathrm{out, in}$, with the domains
 $H^{1}_{0}(\Lambda^j_{a})$ and $H^{1}(\Lambda^j_{a})$
 corresponding to the $\pm$ sign, respectively. Let
 $L^{\pm}_{B,a,\beta}$, $E^{ \mathrm{out},\pm}_{B,a}$, and
 $E^{\mathrm{in},\pm}_{B,a}$ be the self-adjoint operators
 associated with the forms $q^{\pm}_{B,a,\beta}$, $e^{
 \mathrm{out},\pm}_{B,a}$, and $e^{ \mathrm{in},\pm}_{B,a}$,
 respectively.

As in \cite{EY} we are going to use the  Dirichlet-Neumann
 bracketing with additional boundary conditions at the boundary of
 $\Sigma_a$. It works in the magnetic case too as one can see
 easily comparing the form domains of the involved operators --
 cf.~\cite[Thm.~XIII.2]{RS}. We get
 % ------------- %%
 \begin{equation} \label{DNest}
 E^{ \mathrm{in},-}_{B,a}\oplus L^{-}_{B,a,\beta}\oplus E^{
\mathrm{out},-}_{B,a}\leq H_{B,\beta}\leq E^{ \mathrm{in},+}_{B,a}
\oplus L^{+}_{B,a,\beta}\oplus E^{ \mathrm{out},+}_{B,a}
 \end{equation}
 % ------------- %%
with the decomposed estimating operators in $L^2(\mathbb{R}^2)=
L^{2}(\Lambda^{ \mathrm{in}}_{a})\oplus L^{2}(\Sigma_{a})\oplus
L^{2}(\Lambda^{ \mathrm{out}}_{a})$. In order to assess the
 negative eigenvalues of $H_{B,\beta}$, it suffices to consider
 those of $L^{+}_{B,a,\beta}$ and $L^{-}_{B,a,\beta}$, because the
 other operators involved in (\ref{DNest}) are positive. Since the
 loop is smooth, we can pass inside $\Sigma_a$ to the natural
 curvilinear coordinates: we put
 % ------------- %%
 $$ %\begin{equation} \label{}
 (U_{a}f)(s,u)=(1+u\gamma(s))^{1/2}f(\Psi_{a}(s,u))
\quad\mathrm{for}\quad f\in L^{2}(\Sigma_{a})
 $$ %\end{equation}
 % ------------- %%
which defines the unitary operator $U_{a}$ from
 $L^{2}(\Sigma_{a})$ to $L^{2}((0,L)\times (-a,a))$. To express the
 estimating operators in the new variables, we introduce
 % ------------- %%
 \begin{eqnarray*} %\label{}
 Q_{a}^{+} &\!=\!& \Big\{\, \varphi\in H^{1}((0,L)\times
(-a,a));\quad \varphi(L,\cdot)=\varphi(0,\cdot) \quad
\mathrm{on}\quad (-a,a),\\ && \phantom{aAAAAAAAAAAAAAA}
\varphi(\cdot ,a)=\varphi(\cdot, -a)=0 \quad\mathrm{on}\quad (0,L)
\Big \}, \\ Q_{a}^{-} &\!=\!& \Big\{\,\varphi\in H^{1}((0,L)\times
(-a,a)); \quad \varphi(L,\cdot)=
\varphi(0,\cdot)\quad\mathrm{on}\quad (-a,a) \Big\},
 \end{eqnarray*}
 % ------------- %%
and define the quadratic forms
 % ------------- %%
 \begin{eqnarray} \label{estform}
\lefteqn{ b^{\pm}_{B,a,\beta}[g]} \\ &&
=\int^{L}_{0}\int^{a}_{-a}(1+u\gamma(s))^{-2} \left|\partial_s
g\right|^{2}\,du\,ds +\int^{L}_{0}\int^{a}_{-a} \left|\partial_u
g\right|^{2}\,du\, ds \nonumber \\ &&
+\int^{L}_{0}\int^{a}_{-a}V(s,u)|g|^{2}\,ds\,du
-\beta\int^{L}_{0}|g(s,0)|^{2}\,ds \nonumber \\ &&
-\frac{b_{\pm}}{2}\int^{L}_{0}
\frac{\gamma(s)}{1+a\gamma(s)}|g(s,a)|^{2}\,ds
+\frac{b_{\pm}}{2}\int^{L}_{0}
\frac{\gamma(s)}{1-a\gamma(s)}|g(s,-a)|^{2}\,ds \nonumber \\ && +
{1\over 4} \int^{L}_{0}\int^{a}_{-a}
B^{2}(\Gamma_{1}^{2}-2u\Gamma_{1}\Gamma_{2}^{\prime}
+\Gamma_{2}^{2}+2u\Gamma_{2}\Gamma_{1}^{\prime}+u^{2})|g|^{2}\,du\,ds
\nonumber \\ && +B\,{\mathrm{Im}}\int^{L}_{0}\int^{a}_{-a}
(\Gamma_{2}+u\Gamma_{1}^{\prime}) \left((1+u\gamma)^{-1}\cos
H\,\overline{g}\partial_s g -\sin H\,\overline{g}\partial_u g
\right)\,du\,ds \nonumber \\ &&
-B\,{\mathrm{Im}}\int^{L}_{0}\int^{a}_{-a}
(\Gamma_{1}-u\Gamma_{2}^{\prime}) \left( (1+u\gamma)^{-1}\sin H\,
\overline{g}\partial_s g +\cos H\,\overline{g}\partial_u
g\right)\,du\,ds \nonumber
 \end{eqnarray}
 % ------------- %%
on $Q^{\pm}_{a}$, respectively, where $b_+=0$ and $b_-=1$, and
 % ------------- %%
 $$ %\begin{equation} \label{}
V(s,u)= \frac{1}{2}(1+u\gamma(s))^{-3}u\gamma^{\prime\prime}(s)
-\frac{5}{4}(1+u\gamma(s))^{-4}u^{2}\gamma^{\prime}(s)^{2}
-\frac{1}{4}(1+u\gamma(s))^{-2}\gamma(s)^{2}
 $$ %\end{equation}
 % ------------- %%
is the well-known curvature-induced effective potential \cite{ES}.
 Let $D^{\pm}_{B,a,\beta}$ be the self-adjoint operators associated
 with the forms $b^{\pm}_{B,a,\beta}$, respectively. In analogy
 with \cite[Lemma 2.2]{EY}, we get the following result.
 % ------------- %
 \begin{lemma} \label{curvilin}
$\:U^{*}_{a}D^{\pm}_{B,a,\beta}U_{a}=L^{\pm}_{B,a,\beta}.$
 \end{lemma}
 % ------------- %
The presence of the magnetic field gave rise to terms containing
 $\overline{g}\partial_s g$ and $\overline{g}\partial_u g$ in
 (\ref{estform}). In order to eliminate the corresponding
 coefficients modulo small errors, we employ another unitary
 operator. We put
 % ------------- %%
 $$ %\begin{equation} \label{}
T_{B}(s)=-{1\over 2} B\int^{s}_{0}
\left(\Gamma_{2}(t)\Gamma_{1}^{\prime}(t)
-\Gamma^{\prime}_{2}(t)\Gamma_{1}(t)\right)\,dt\,;
 $$ %\end{equation}
 % ------------- %%
it follows from the Green theorem that $T_{B} (L)=B|\Omega|\,$.
 Then we define
 % ------------- %%
 $$ %\begin{equation} \label{}
(M_{B}h)(s,u):=\exp\Big\lbrack iT_{B}(s)+{i\over 2}Bu
\left(\Gamma_{2}(s)\sin H(s)+\Gamma_{1}(s)\cos H(s)\right)
\Big\rbrack\,h(s,u)
 $$ %\end{equation}
 % ------------- %%
for any $h\in L^{2}((0,L)\times (-a,a))$; it is straightforward to
 check that $M_{B}$ is a unitary operator on $L^{2}((0,L)\times
(-a,a))$. We define
 % ------------- %%
 \begin{eqnarray*} %\label{}
 \tilde{Q}_{B,a}^{+} &\!=\!& \Big\{\, \varphi\in H^{1}((0,L)\times
(-a,a));\quad \varphi(L,\cdot)=
\mathrm{e}^{-iB|\Omega|}\varphi(0,\cdot) \; \mathrm{on}\;
(-a,a),\\ && \phantom{aAAAAAAAAAAAAAAAA} \varphi(\cdot
,a)=\varphi(\cdot, -a)=0 \quad\mathrm{on}\quad (0,L) \Big \}, \\
\tilde{Q}_{B,a}^{-} &\!=\!& \Big\{\,\varphi\in H^{1}((0,L)\times
(-a,a)); \quad \varphi(L,\cdot)=\mathrm{e}^{-iB|\Omega|}
\varphi(0,\cdot) \;\mathrm{on}\;(-a,a) \Big\},
 \end{eqnarray*}
 % ------------- %%
and another pair of quadratic forms
 % ------------- %%
 \begin{eqnarray*} %\label{}
\lefteqn{ \tilde{b}^{\pm}_{B,a,\beta}[g]} \\ &&
=\int^{L}_{0}\int^{a}_{-a} \Big\{ (1+u\gamma)^{-2}|\partial_s
g|^{2} +|\partial_u g|^{2} \\ && + \big\lbrack
B(\Gamma_{2}+u\Gamma_{1}^{\prime})(1+u\gamma)^{-1}\cos H
-B(\Gamma_{1}-u\Gamma_{2}^{\prime})(1+u\gamma)^{-1}\sin H \\ &&
-B(1+u\gamma)^{-2}(\Gamma_{2}\cos H-\Gamma_{1}\sin H) \\ && +
B(1+u\gamma)^{-2}(\Gamma_{2}\sin H+\Gamma_{1}\cos H)^{\prime}u
\big\rbrack\, {\mathrm{Im}}(\overline{g}\partial_s g)
+W_{B}(s,u)|g|^{2} \Big\}\,du\,ds \\ &&
-\beta\int^{L}_{0}|g(s,0)|^{2}\,ds
-\frac{b_{\pm}}{2}\int^{L}_{0}\frac{\gamma(s)}{1+a\gamma(s)}
|g(s,a)|^{2}\,ds \\ &&
+\frac{b_{\pm}}{2}\int^{L}_{0}\frac{\gamma(s)}
{1-a\gamma(s)}|g(s,-a)|^{2}\,ds
 \end{eqnarray*}
 % ------------- %%
for $g\in\tilde{Q}^{\pm}_{B,a}$, respectively, where
 % ------------- %%
 \begin{eqnarray*} %\label{}
\lefteqn{ W_{B}(s,u)} \\ && =V(s,u)+ {1\over 4}
(1+u\gamma)^{-2}B^{2}u^{2}((\Gamma_{2}\sin H+\Gamma_{1} \cos
H)^{\prime})^{2} \\ && + {1\over 4}
B^{2}(\Gamma_{1}^{2}-2u\Gamma_{1}\Gamma_{2}^{\prime}
+\Gamma_{2}^{2}+2u\Gamma_{2}\Gamma_{1}^{\prime}+u^{2}) \\ &&
+B(\Gamma_{2}\!+\!u\Gamma_{1}^{\prime})(1+u\gamma)^{-1}
T_{B}^{\prime}(s)\cos H
-B(\Gamma_{1}\!-\!u\Gamma_{2}^{\prime})(1+u\gamma)^{-1}
T_{B}^{\prime}(s)\sin H \\ && + {1\over 4}
(1+u\gamma)^{-2}B^{2}(\Gamma_{2}\cos H-\Gamma_{1}\sin H)^{2} +
{1\over 4} B^{2}(\Gamma_{2}\sin H+\Gamma_{1}\cos H)^{2} \\ &&
+[B(\Gamma_{2}+u\Gamma_{1}^{\prime})(1+u\gamma)^{-1}\cos H
-B(\Gamma_{1}-u\Gamma_{2}^{\prime})(1+u\gamma)^{-1}\sin H \\ &&
-B(1+u\gamma)^{-2}(\Gamma_{2}\cos H-\Gamma_{1}\sin H)]\, {1\over
2} B(\Gamma_{2}\sin H+\Gamma_{1}\cos H)^{\prime}u \\ &&
+[-B(\Gamma_{2}+u\Gamma_{1}^{\prime})\sin H
-B(\Gamma_{1}-u\Gamma_{2}^{\prime})\cos H]\, {1\over 2}B
(\Gamma_{2}\sin H+\Gamma_{1}\cos H)\,.
 \end{eqnarray*}
 % ------------- %%
Let $\tilde{D}^{\pm}_{B,a,\beta}$ be the self-adjoint operators
 associated with the forms $\tilde{b}^{\pm}_{B,a,\beta}$,
 respectively. By a straightforward computation, one can check the
 following claim.
 % ------------- %
 \begin{lemma} \label{gaugeout}
 $\:M_{B}^{*}D^{\pm}_{B,a,\beta} M_{B}=\tilde{D}^{\pm}_{B,a,\beta}\,.$
 \end{lemma}
 % ------------- %
The next step is to estimate $\tilde{D}^{\pm}_{B,a,\beta}$ by
 operators with separated variables. Denoting
 % ------------- %%
 $$ %\begin{equation} \label{}
\gamma_{+:}=\max_{[0,L]}|\gamma(\cdot)|
 $$ %\end{equation}
 % ------------- %%
we put
 % ------------- %%
 \begin{eqnarray*} %\label{}
N_{B}(a) &\!:=\!& \max_{(s,u)\in [0,L]\times [-a,a]} \Big|
B(\Gamma_{2}+u\Gamma_{1}^{\prime})(1+u\gamma)^{-1}\cos H \\ &&
-B(\Gamma_{1}-u\Gamma_{2}^{\prime})(1+u\gamma)^{-1}\sin H \\ &&
-B(1+u\gamma)^{-2}(\Gamma_{2}\cos H-\Gamma_{1}\sin H) \\ &&
+B(1+u\gamma)^{-2}(\Gamma_{2}\sin H+\Gamma_{1}\cos H)^{\prime}u
\Big|
 \end{eqnarray*}
 % ------------- %%
and
 % ------------- %%
 $$ %\begin{equation} \label{}
M_{B}(a) :=\max_{(s,u)\in [0,L]\times [-a,a]} \left|
W_{B}(s,u)+\frac{1}{4}\gamma(s)^{2} \right|\,.
 $$ %\end{equation}
 % ------------- %%
Let $\emptyset\neq I\subset\mathbb{R}$ be a compact interval. Then
 there is a positive $K$ such that
 % ------------- %%
 $$ %\begin{equation} \label{}
N_{B}(a)+M_{B}(a)\leq Ka \quad\mathrm{for}\quad
0<a<\frac{1}{2\gamma_{+}}\quad\mathrm{and} \quad B\in I\,,
 $$ %\end{equation}
 % ------------- %%
where $K$ is independent of $a$ and $B$. For a fixed
 $0<a<\frac{1}{ 2\gamma_{+}}$, we define
 % ------------- %%
 \begin{eqnarray*} %\label{}
\hat{b}^{\pm}_{B,a,\beta}[f] &\!:=\! &\int^{L}_{0}\int^{a}_{-a}
\Bigg\{ \left\lbrack(1 \mp a\gamma_{+})^{-2} \pm
\frac{1}{2}N_{B}(a) \right\rbrack |\partial_s f|^{2}+|\partial_u
f|^{2}
\\ && + \left\lbrack -\frac{1}{4}\gamma(s)^{2} \pm \frac{1}{2}N_{B}(a)
\pm M_{B}(a) \right\rbrack|f|^{2} \Bigg\}\,du\,ds \\ &&
-\beta\int^{L}_{0}|f(s,0)|^{2}\,ds -\gamma_{+}b_{\pm}\int^{L}_{0}
(|f(s,a)|^{2}+|f(s,-a)|^{2})\,ds
 \end{eqnarray*}
 % ------------- %%
for $f\in\tilde{Q} ^{\pm}_{B,a}$, respectively. Since
 $|{\mathrm{Im}}(\overline{g}\partial_s g)| \leq
\frac{1}{2}(|g|^{2}+|\partial_s g|^{2})$, we obtain
 % ------------- %%
 \begin{eqnarray}
\tilde{b}^{+}_{B,a,\beta}[f] &\!\leq\!&
\hat{b}^{+}_{B,a,\beta}[f]\quad\mathrm{for}\quad
f\in\tilde{Q}^{+}_{B,a}\,, \label{upper} \\
\hat{b}^{-}_{B,a,\beta}[f] &\!\leq\!&
\tilde{b}^{-}_{B,a,\beta}[f]\quad\mathrm{for}\quad f\in
\tilde{Q}^{-}_{B,a}\,.\label{lower}
 \end{eqnarray}
 % ------------- %%
Let $\hat{H}^{\pm}_{B,a,\beta}$ be the self-adjoint operators
 associated with the forms $\hat{b} ^{\pm}_{B,a,\beta}$,
 respectively. Furthermore, let $T^{+}_{a,\beta}$ be the
 self-adjoint operator associated with the form
 % ------------- %%
 $$ %\begin{equation} \label{}
t^{+}_{a,\beta}[f]=
\int^{a}_{-a}|f^{\prime}(u)|^{2}\,ds-\beta|f(0)|^{2},\quad f\in
H_0^{1}((-a,a))\,,
 $$ %\end{equation}
 % ------------- %%
and similarly, let $T^{-}_{a,\beta}$ be the self-adjoint operator
 associated with the form
 % ------------- %%
 $$ %\begin{equation} \label{}
t^{-}_{a,\beta}[f]=
\int^{a}_{-a}|f^{\prime}(u)|^{2}\,ds-\beta|f(0)|^{2}
-\gamma_{+}(|f(a)|^{2}+|f(-a)|^{2}),\quad f\in H^{1}((-a,a))\,.
 $$ %\end{equation}
 % ------------- %%
We define
 % ------------- %%
 $$ %\begin{equation} \label{}
U_{B,a}^{\pm}= -\left\lbrack(1\mp a\gamma_{+})^{-2} \pm
\frac{1}{2}N_{B}(a) \right\rbrack\frac{d^{2}}{ds^{2}}
-\frac{1}{4}\gamma(s)^{2} \pm\frac{1}{2}N_{B}(a) \pm M_{B}(a)
 $$ %\end{equation}
 % ------------- %%
in $L^{2}((0,L))$ with the domain $P_{B}$ specified in the
 previous section. Then we have
 % ------------- %%
 \begin{equation} \label{decomp}
\hat{H}^{\pm}_{B,a,\beta}= U_{B,a}^{\pm}\otimes 1+1\otimes
T^{\pm}_{a,\beta}.
 \end{equation}
 % ------------- %%
Let $\mu_{j}^{\pm}(B,a)$ be the $j$-th eigenvalue of
 $U^{\pm}_{B,a}$ counted with multiplicity. We shall prove the
 following estimate.
 % ------------- %%
 \begin{proposition} \label{evest}
 Let $j\in\mathbb{N}$. Then there exists $C(j)>0$ such that
 % ------------- %%
 $$ %\begin{equation} \label{}
|\mu_{j}^{+}(B,a)-\mu_{j}(B)|+ |\mu_{j}^{-}(B,a)-\mu_{j}(B)|\leq
C(j)a
 $$ %\end{equation}
 % ------------- %%
holds for $B\in I$ and $0<a<\frac{1}{2\gamma_{+}}$, where $C(j)$
 is independent of $B$ and $a$.
 \end{proposition}
 % ------------- %%
{\sl Proof:} Since
 % ------------- %%
 \begin{eqnarray*} %\label{}
\lefteqn{ U^{+}_{B,a} -\left\lbrack
(1-a\gamma_{+})^{-2}+\frac{1}{2}N_{B}(a) \right\rbrack S_{B}} \\
&& = \frac{1}{4}\left\lbrack
\frac{a\gamma_{+}(2-a\gamma_{+})}{(1-a\gamma_{+})^{2}}
+\frac{1}{2}N_{B}(a) \right\rbrack
\gamma(s)^{2}+\frac{1}{2}N_{B}(a)+M_{B}(a)\,,
 \end{eqnarray*}
 % ------------- %%
and since $N_{B}(a)+M_{B}(a)\leq Ka$ for $0<a<
\frac{1}{2\gamma_{+}}$ and $B\in I$, we infer that there is a
 constant $C_{1}>0$ such that
 % ------------- %%
 $$ %\begin{equation} \label{}
\left\Vert U^{+}_{B,a} - \left\lbrack
(1-a\gamma_{+})^{-2}+\frac{1}{2}N_{B}(a) \right\rbrack S_{B}
\right\Vert \leq C_{1}a
 $$ %\end{equation}
 % ------------- %%
for $0<a<\frac{1}{2\gamma_{+}}$ and $B\in I$. This together with
 the min-max principle implies that
 % ------------- %%
 $$ %\begin{equation} \label{}
\left|\mu^{+}_{j}(B,a)- \left\lbrack
(1-a\gamma_{+})^{-2}+\frac{1}{2}N_{B}(a) \right\rbrack \mu_{j}(B)
\right | \leq C_{1}a
 $$ %\end{equation}
 % ------------- %%
for $0<a<\frac{1}{2\gamma_{+}}$ and $B\in I$. Since
 $\mu_{j}(\cdot)$ is continuous, we claim that there exists a
 constant $C_{2}>0$ such that
 % ------------- %%
 $$ %\begin{equation} \label{}
\left|\mu^{+}_{j}(B,a)-\mu_{j}(B)\right| \leq C_{2}a
 $$ %\end{equation}
 % ------------- %%
for $0<a<\frac{1}{2\gamma_{+}}$ and $B\in I$. In a similar way, we
 infer the existence of a constant $C_{3}>0$ such that
 % ------------- %%
 $$ %\begin{equation} \label{}
\left|\mu^{-}_{j}(B,a)-\mu_{j}(B)\right| \leq C_{3}a
 $$ %\end{equation}
 % ------------- %%
for $0<a<\frac{1}{2\gamma_{+}}$ and $B\in I$. \QED \vspace{1.2em}

We also recall the following result from [EY].
 % ------------- %%
 \begin{proposition} \label{propEY}
(a) Suppose that $\beta a>\frac{8}{3}$. Then $T^{+}_{a,\beta}$ has
 only one negative eigenvalue, which we denote by
 $\zeta^{+}_{a,\beta}$. It satisfies the inequalities
 % ------------- %%
 $$ %\begin{equation} \label{}
-\frac{1}{4}\beta^{2}<\zeta^{+}_{a,\beta}<
-\frac{1}{4}\beta^{2}+2\beta^{2}\exp \left(-\frac{1}{2}\beta
a\right)\,.
 $$ %\end{equation}
 % ------------- %%
(b) Let $a\beta>8$ and $\beta>\frac{8}{3}\gamma_{+}$. Then
 $T^{-}_{a,\beta}$ has a unique negative eigenvalue
 $\zeta^{-}_{a,\beta}$, and moreover, we have
 % ------------- %%
 $$ %\begin{equation} \label{}
-\frac{1}{4}\beta^{2}-
\frac{2205}{16}\beta^{2}\exp\left(-\frac{1}{2}\beta a\right)
<\zeta^{-}_{a,\beta}<-\frac{1}{4}\beta^{2}\,.
 $$ %\end{equation}
 % ------------- %%
\end{proposition}
% ------------- %%
\vspace{1.2em}

\noindent Now we are ready to prove Theorem~\ref{main}. We put
 $a(\beta) =6\beta^{-1}\ln\beta$. Let $\xi^{\pm}_{\beta,j}$ be
 the $j$-th eigenvalue of $T^{\pm}_{a(\beta),\beta}$, by
 Proposition~\ref{propEY} we have
 % ------------- %%
 $$ %\begin{equation} \label{}
\xi^{\pm}_{\beta,1}=\zeta^{\pm}_{a(\beta)\,,\beta},\quad
\xi_{\beta,2}^{\pm}\geq 0\,.
 $$ %\end{equation}
 % ------------- %%
>From the decompositions (\ref{decomp}) we infer that $\{
\xi^{\pm}_{\beta,j} +\mu_{k}^{\pm}
(B,a(\beta))\}_{j,k\in{\mathbb{N}}}$, properly ordered, is the
 sequence of the eigenvalues of $\hat{H}^{\pm}_{ B,a(\beta),\beta}$
 counted with multiplicity. Proposition~\ref{evest} gives
 % ------------- %%
 \begin{equation} \label{lowest}
\xi^{\pm}_{\beta,j}+\mu_{k}( B,a(\beta))\geq \mu_{1}^{\pm}(
B,a(\beta))=\mu_{1}(B)+\mathcal{O}(\beta^{-1}\ln\beta)
 \end{equation}
 % ------------- %%
for $B\in I$, $j\geq 2$, and $k\geq 1$, where the error term is
 uniform with respect to $B\in I$. For a fixed $j\in\mathbb{N}$, we
 put
 % ------------- %%
 $$ %\begin{equation} \label{}
\tau^{\pm}_{B,\beta,j}=\zeta^{\pm}_{a(\beta),\beta}
+\mu_{j}^{\pm}(B,a(\beta)).
 $$ %\end{equation}
 % ------------- %%
Combining Propositions~\ref{evest} and \ref{propEY} we get
 % ------------- %%
 \begin{equation} \label{tauasympt}
\tau^{\pm}_{B,\beta,j}=-\frac{1}{4}\beta^{2}
+\mu_{j}(B)+\mathcal{O}(\beta^{-1}\ln\beta)\quad\mathrm{as}
\quad\beta \to\infty\,,
 \end{equation}
 % ------------- %%
where the error term is uniform with respect to $B\in I$. Let us
 fix now $n\in\mathbb{N}$. Combining (\ref{lowest}) with
 (\ref{tauasympt}) we infer that there exists $\beta(n,I)>0$
 such that the inequalities
 % ------------- %%
 $$ %\begin{equation} \label{}
\tau^{+}_{B,\beta,n}<0,\quad
\tau^{+}_{B,\beta,n}<\xi^{+}_{\beta,j}+\mu_{k}^{+}(
B,a(\beta)),\quad
\tau^{-}_{B,\beta,n}<\xi^{-}_{\beta,j}+\mu_{k}^{-}( B,a(\beta))
 $$ %\end{equation}
 % ------------- %%
hold for $B\in I$, $\beta\geq\beta(n,I)$, $j\geq 2$, and $k\geq
1$. Hence the $j$-th eigenvalue of $\hat{H}^{\pm}_{ B,a(\beta),
\beta}$ counted with multiplicity is $\tau^{\pm}_{B,\beta,j}$
 for $B\in I$, $j\leq n$, and $\beta\geq \beta(n,I)$. Let $B\in
I$ and $\beta\geq \beta(n,I)$. We denote by  $\kappa^{\pm}_{j}
(B,\beta)$ the $j$-th eigenvalue of $L^{\pm}_{B,a,\beta}$.
 Combining our basic estimate (\ref{DNest}) with
 Lemmas~\ref{curvilin} and \ref{gaugeout}, relations
 (\ref{upper}) and (\ref{lower}), and the min-max principle, we
 arrive at the inequalities
 % ------------- %%
 \begin{equation} \label{finalest}
\tau^{-}_{B,\beta,j}\leq\kappa^{-}_{j}(B,\beta)\quad
\mathrm{and}\quad \kappa^{+}_{j}(B,\beta)\leq\tau^{+}_{
B,\beta,j}\quad\mathrm{for}\quad 1\leq j\leq n\,,
 \end{equation}
 % ------------- %%
so we have $\kappa^{+}_{n}(B,\beta)<0<\inf
\sigma_{\mathrm{ess}}(H_{B,\beta})$. Hence the min-max principle
 and (\ref{DNest}) imply that $H_{B,\beta}$ has at least $n$
 eigenvalues in $(-\infty,\kappa^{+}_{n}(B,\beta)]$. Given $1\leq
j\leq n$, we denote by $\lambda_{j}(B,\beta)$ the $j$-th
 eigenvalue of $H_{B,\beta}$. It satisfies
 % ------------- %%
 %\begin{equation}
$$\kappa_{j}^{-}(B,\beta)\leq\lambda_{j} (B,\beta)
\leq\kappa_{j}^{+} (B,\beta)\quad\mathrm{for}\quad 1\leq j\leq
n\,;$$
% \end{equation}
 % ------------- %%
this together with (\ref{tauasympt}) and (\ref{finalest})
 implies that
 % ------------- %%
 $$ %\begin{equation} \label{}
\lambda_{j}(B,\beta)=-\frac{1}{4}\beta^{2}
+\mu_{j}(B)+\mathcal{O}(\beta^{-1}\ln\beta)
\quad\mathrm{as}\quad\beta \to\infty\quad \mathrm{for} \quad
1\leq j\leq n\,,
 $$ %\end{equation}
 % ------------- %%
where the error term is uniform with respect to $B\in I$. This
 completes the proof. \QED \vspace{1.2em}

\noindent{\sl Proof of Corollary~\ref{nonconst}}: By
\cite[Thm~XIII.89]{RS} no eigenvalue $\mu_{n}(\cdot)$ is
constant on $I$. This together with Theorem~\ref{main} yields
the claim. \QED

%%%%%%%%%%%%%%%%%%%%%%%%%%%%%%%%%%%%%%%%%%%%%%%%%%%%%%%%%%%%%%%%%%%

\setcounter{equation}{0}
\section{Concluding remarks}

The above corollary answers the question we posed in the
 introduction as a mathematical problem showing that a ring with a
 strong enough attractive $\delta$-interaction does exhibit
 persistent currents. On the other hand, from the physical point of
 view it would be bold to identify a mere non-constantness of the
 eigenvalues with a genuine magnetic transport around the loop.

The problem is similar to other situation where an electron can be
 transported in a magnetic field due to the presence of a
 ``guiding'' perturbation. A prime example are the edge currents
 \cite{Ha, MS} which attracted a wave of mathematical interest
 recently in connection with the problem of stability of the
 transport with respect to perturbations. In case of a single edge
 and a weak disorder a part of the absolutely continuous
 spectrum survives \cite{BP, FGW, MMP} but the fact itself gives no
 quantitative information about the transport. On the other hand, a
 system with more than one edge may have no continuous spectrum at
 all and still it has states in which electrons travel distances
 much larger than the corresponding cyclotron radius \cite{FM}.

In our case it is clear, for instance, that the loop geometry
 influences the transport substantially. If $\Gamma$ is a circle,
 e.g., than up the $\mathcal{O}(\beta^{-1}\ln\beta)$ error the
 persistent-current plot will have the ideal saw-tooth shape as we
 can see from the relation (\ref{ideal}); one expects that the
 eigenfunctions will be ``spread'' around the whole circle. In
 contrast, if the loop is rather ``wiggly'' the one-dimensional
 comparison operator $S_B$ contains an irregular effective
 potential coming from the rapidly varying curvature, which may
 cause -- depending on the strength of such a ``disorder'' -- that
 the most part of the electron wavefunction will be concentrated in
 (the vicinity of) a part of the loop only. The same may happen if
 the loop curvature is slowly changing but a disorder potential is
 added to the Hamiltonian.

To distinguish the situations with a {\em significant} transport,
 one needs clear\-ly to understand better the sketched
 ``disordered'' cases which do not fall into this category. We
 leave this problem to a future publication.

%%%%%%%%%%%%%%%%%%%%%%%%%%%%%%%%%%%%%%%%%%%%%%%%%%%%%%%%%%%%%%%%%%%

\subsection*{Acknowledments}

The research has been partially supported by GAAS and the Czech
 Ministry of Education within the projects A1048101 and ME170. %

\end{document}